\documentclass{aa}
\usepackage{graphicx}
\def\ltsim{\raise 2pt \hbox {$<$} \kern-1.1em \lower 4pt \hbox {$\sim$}}
\def\ltapprox{\raise 2pt \hbox {$<$} \kern-1.1em \lower 5pt \hbox {$\approx$}}
\def\gtsim{\raise 2pt \hbox {$>$} \kern-1.1em \lower 4pt \hbox {$\sim$}}
\def\gtapprox{\raise 2pt \hbox {$>$} \kern-1.1em \lower 5pt \hbox {$\approx$}}

\def\arcsec{$^{\prime\prime}$}
\def\arcmin{$^{\prime}$}
\def\degrees{$^{\circ}$}
\def\etal{{et al.~}}
\begin{document}
%
\title
{Spectral index maps of the radio halos \\in Abell 665 and Abell 2163 }
\author{L. Feretti\inst{1} 
\and E. Orr\`u\inst{1} 
\and G. Brunetti\inst{1} 
\and G. Giovannini\inst{1,2} 
\and N. Kassim\inst{3}
\and G. Setti\inst {1,2}}
\offprints{L. Feretti}
\institute{
Istituto di Radioastronomia -- CNR, via P. Gobetti 101, I--40129
Bologna, Italy
\and Dipartimento di Astronomia, Univ. Bologna, via Ranzani 1, 
I-40127 Bologna, Italy
\and Naval Research Laboratory, Code 7213, Washington, DC, 20375 USA }

   \date{Submit to Astronomy \& Astrophysics}

\abstract{
New radio data at 330 MHz are presented for the rich clusters Abell
665 and Abell 2163, whose radio emission is characterized by the
presence of a radio halo.  These images allowed us to derive the
spectral properties of the two clusters under study.  The integrated
spectra of these halos between 0.3 GHz and 1.4 GHz are moderately
steep: $\alpha^{1.4}_{0.3}$ = 1.04 and $\alpha^{1.4}_{0.3}$ = 1.18,
for A665 and A2163, respectively.  The spectral index maps, produced
with an angular resolution of the order of $\sim$ 1\arcmin, show
features of the spectral index (flattening and patches), which are
indication of a complex shape of the radiating electron spectrum, and
are therefore in support of electron reacceleration models.  Regions
of flatter spectrum are found to be related to the recent merger
activity in these clusters.  This is the first strong confirmation
that the cluster merger supplies energy to the radio halo. In the
undisturbed cluster regions, the spectrum steepens with the distance
from the cluster center.  This is interpreted as the result of the
combination of the magnetic field profile with the spatial
distribution of the reacceleration efficiency, thus allowing us to set
constraints on the radial profile of the cluster magnetic field.
}

\maketitle

\section{Introduction}

Recent observations of clusters of galaxies have revealed a new and
complex scenario in the structure of the intergalactic medium.  The
clusters are not simple relaxed structures, but are still forming at
the present epoch.  Substructures, commonly observed in the X-ray
distribution of a high number of rich clusters (Henry \& Briel 1993,
Burns \etal 1994), are evidence of the hierarchic growth of clusters
from the merger of poorer subclusters.

\begin{table*}
\caption{Observed clusters of galaxies}
\begin{flushleft}
\begin{tabular}{lllllllllll}
\hline
\noalign{\smallskip}
Name  & z  &  RA (J2000) & DEC  & T & L$_X$ (bol) & r$_c$ & r$_c$ & $\beta$ &
P$_{1.4}$ & LS \\
      &    &  h~~ m~~ s  & \degrees~~~~ \arcmin~~~~ \arcsec~~ 
& keV & erg s$^{-1}$ & \arcsec  &  kpc & & W Hz$^{-1}$ & Mpc\\
\noalign{\smallskip}
\hline
\noalign{\smallskip}
A~665  & 0.1818 &  08 30 57.4 & +65 51 14.4  & 9.03  & 4.17 10$^{45}$ & 
96 & 379 & 0.66  & 6.6 10$^{24}$  & 1.8  \\
A~2163 & 0.203  &  16 15 49.4 & --06 09 00.0  & 13.83 & 1.33 10$^{46}$ & 
72 & 305 & 0.62  & 3.0 10$^{25}$  & 2.9  \\
\noalign{\smallskip}
\hline
\label{olog}
\end{tabular}
\end{flushleft}
\par\noindent
Caption. Col. 1: cluster name; Col. 2: redshift;
Cols. 3 and 4: coordinates of the X-ray cluster center from Ebeling et al.
(1996);
Col. 5: temperature from Allen \& Fabian (1995); 
Col. 6: bolometric X-ray luminosity from Wu et al. (1999);
Cols. 7, 8 and 9: cluster core radius (angular and linear) and $\beta$ from
Birkinshaw et al. (1991)  for A665 and Elbaz et al. (1995) form A2163 ; 
Cols. 10 and 11: Total radio power 
and maximum linear size at 1.4 GHz from Giovannini \& Feretti (2000) 
for A665 and Feretti et al. (2001) for A2163.
\end{table*}

An important problem in cluster phenomenology involves cluster-wide
radio halos, whose prototype is Coma C (Giovannini \etal 1993, and
references therein). These are extended diffuse radio sources located
at the cluster centers, with typical sizes of \gtsim 1 $h_{50}^{-1}$
Mpc, regular shape, steep radio spectra and no significant
polarization.  According to recent suggestions, the cluster merger
process may play a crucial role in the formation and energetics of
these sources (see Giovannini \& Feretti 2002, and references
therein).

The origin and evolution of halos is still a matter of debate.
Several suggestions for the mechanism transferring energy into the
relativistic electron population and for the origin of relativistic
electrons themselves have been made: in-situ reacceleration of
relativistic electrons by plasma and by shock waves, particle
injection from radio galaxies, acceleration out of the thermal pool,
secondary electrons resulting from hadronic collisions of relativistic
protons with the ICM gas protons, and combinations of these processes
(see e.g. Brunetti 2003, Blasi 2003, Petrosian 2003).  It is therefore
important to carry out new observations aimed at discriminating
between these theoretical models.  

An important observable in a radio
halo is the spectral index distribution.  It could allow us to test
predictions of these different models, since it reflects the shape of
the electron energy distribution.
Spectral index maps represent a powerful tool to study the properties
of the relativistic electrons and of the magnetic field in which they
emit, and to investigate the connection between the electron energy
distribution and the ICM.  By combining high resolution spectral
information and X--ray images it is possible to study the
thermal--relativistic plasma connection both on small scales
(e.g. spectral index variations vs. clumps in the ICM distribution)
and on the large scale (e.g. radial spectral index trends).

The prototypical example of cluster radio halos is the diffuse source
Coma C in the Coma cluster (Willson 1970).  Nowadays Coma C is the
only radio halo for which a high resolution spectral index image has
been presented in the literature (Giovannini et al., 1993).  The
spectral index trend in Coma C shows a flat spectrum\footnote
{$S_{\nu}\propto \nu ^{-\alpha}$ through this paper} in the center
($\alpha\simeq$ 0.8) and a progressive steepening with increasing
distance from the center (up to $\alpha\simeq$ 1.8). Since the
diffusion velocity of relativistic particles is low in relation to
their radiative lifetime, the radial spectral steepening cannot be
simply due to ageing of radioemitting electrons. Therefore the
spectral steepening must be related to the intrinsic evolution of the
local electron spectrum and to the radial profile of the cluster
magnetic field.

It has been shown (Brunetti et al. 2001) that a relatively general
expectation of models invoking reacceleration of relic particles is a
radial spectral steepening in the synchrotron emission from radio
halos.  The steepening, that is difficult to reproduce by other models
such as those invoking secondary electron populations, is due to the
combined effect of a radial decrease of the cluster magnetic field
strength and of the presence of a high energy break in the energy
distribution of the reaccelerated electron population.  In the
framework of reacceleration models the radio spectral index map can be
used to derive the physical conditions prevailing in the clusters,
i.e. reacceleration efficiency and magnetic field
strength.

This method has been successfully applied to the case of Coma C by
Brunetti et al. (2001) who applied a {\it two phase} reacceleration
model and obtained large scale reacceleration efficiencies of the
order of $\sim 10^{8}$ yr$^{-1}$ and magnetic field strengths ranging
from 1--3 $\mu$G in the central regions down to 0.05--0.1 $\mu$G in the
cluster periphery.

In this paper we present the radio images at 90 cm of Abell 665 and
Abell 2163. These clusters host powerful and giant radio halos studied
at 20 cm with the VLA (Giovannini \& Feretti 2000, Feretti et
al. 2001).  From the comparison between the 20 and 90 cm data, we
derive the spectral index maps of these radio halos and discuss their
implications.

We adopt H$_0$=50 km s$^{-1}$ Mpc$^{-1}$ and q$_0$ = 0.5. With these
values, 1 arcsec corresponds to 3.94 kpc at the distance of A665 and
to 4.27 kpc at the distance of A2163.

\begin{table}
\caption{Observing log}
\begin{flushleft}
\begin{tabular}{llllllll}
\hline
\noalign{\smallskip}
Name & $\nu$ &  $\Delta \nu$  & Conf  &  Date  & Dur \\
  & MHz & MHz  & & & h \\
\noalign{\smallskip}
\hline
\noalign{\smallskip}
A665 & 321.5/327.5 & 3.125 & B & 4/2001 & 3  \\
     &  `` & `` & C & 9/2001 & 6 \\
     & `` & `` & D & 12/2001 & 1.5 \\
A2163 &  321.5/327.5 & 3.125 & B & 4/2001 & 3.5   \\
       &  `` & `` & C & 8/2001 & 6  \\
\noalign{\smallskip}
\hline
\noalign{\smallskip}
\end{tabular}
\end{flushleft}
Caption. Col. 1: cluster name; Col. 2: observing frequencies;
Cols. 3: bandwidth; Col. 4: VLA configuration;
Col. 5: month and year of observation; Col. 6: observing time duration.
\end{table}

\begin{figure*}
\includegraphics{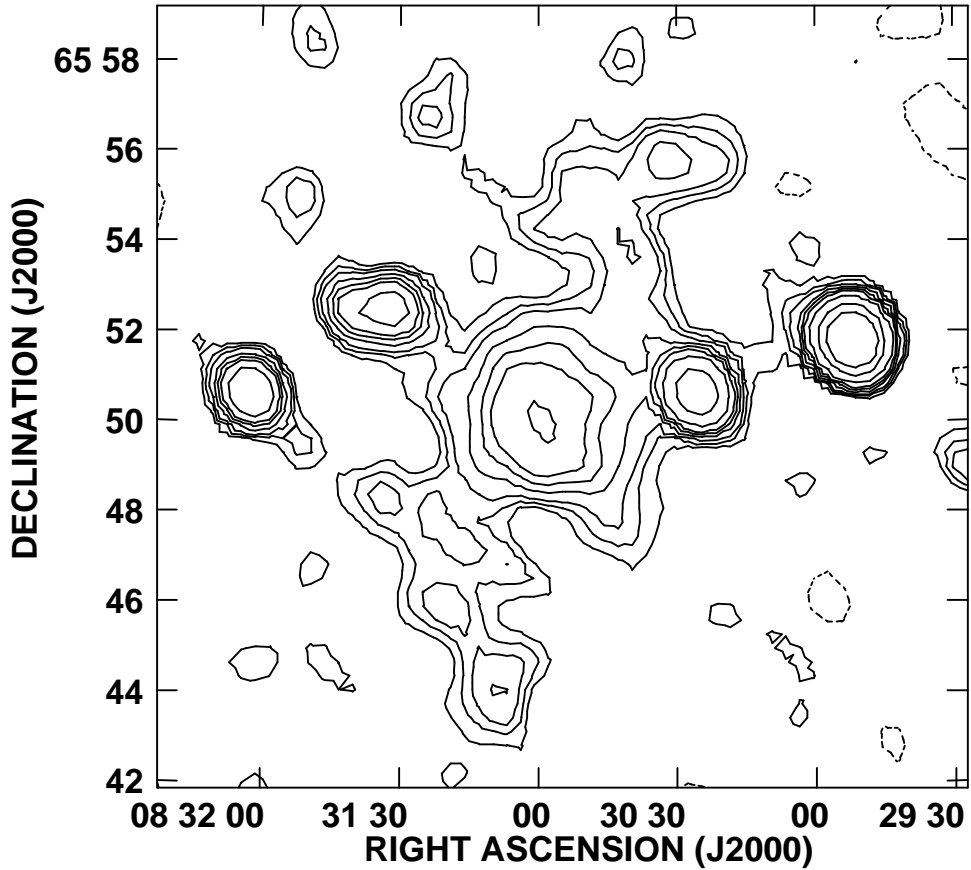}
\vspace{14 cm}
\caption{ Radio map at 90 cm of A665. The beam FWHM is 68\arcsec 
$\times$ 59\arcsec~ in PA= 25\degrees. Contour levels are -2, 2, 3, 4, 6, 8, 
10, 15, 30, 50 mJy/beam.}
\end{figure*}

\begin{figure*}
\includegraphics{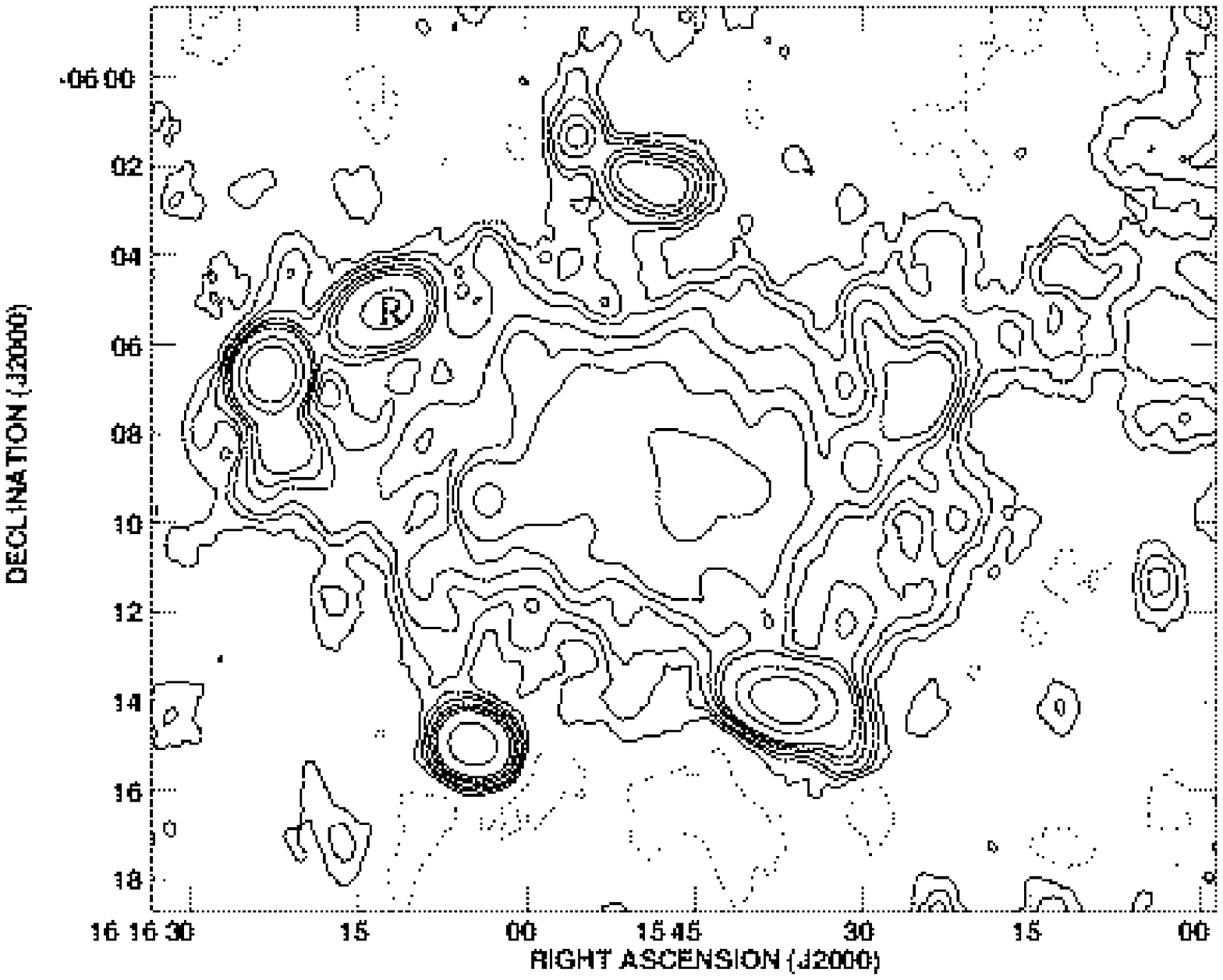}
\vspace{14 cm}
\caption{ Radio map at 90 cm of A2163. The beam FWHM is 57\arcsec 
$\times$
51\arcsec~ in PA= 2\degrees. Contour levels are -1, 1, 2, 3, 5, 7, 
10, 25, 50 mJy/beam.}
\end{figure*}

\section{The clusters}

The general properties of A665 and A2163 are summarized in Table 1.
We give below a short description of the radio and X-ray information.

{\bf A~665}.  The presence of a radio halo was first reported by
Moffet \& Birkinshaw (1989) and confirmed by Jones \& Saunders (1996).
The deep VLA image at 20 cm, presented by Giovannini \& Feretti
(2000), shows that the radio halo is about 1.8 Mpc in size, with
elongation in the SE-NW direction.  The radio emission is asymmetric
with respect to the cluster center, being brighter and more extended
toward NW.

The X-ray emission of this cluster is complex.  An acceptable fit of
the X-ray brightness profile to a $\beta$ model was obtained by
Birkinshaw et al. (1991) using the Einstein IPC data.  In the
following years, however, Hughes \& Birkinshaw (1994) found that the
spatial distribution of the X-ray emitting gas in the ROSAT PSPC data
deviates from circular symmetry and argued that a major merger is
occurring or had recently occurred in this system.  This was confirmed
by G\'omez et al. (2000), who studied the cluster dynamics on the
basis of redshift measurements and concluded that a merger of two
similar mass subclusters seen close to the time of core-crossing,
produces velocity distributions that are consistent with the
observations.

Chandra data presented by Markevitch \& Vikhlinin (2001) indicate that
the X-ray contours are elongated in the same direction as the galaxy
distribution (Geller \& Beers 1982) and their appearance suggests that
the bright core associated with the main galaxy concentration is
moving south with respect to the more diffuse cluster component. This
is confirmed by their temperature map, which reveals a remarkable
shock in front of the cool cluster core, indicating that the core is
moving with a relatively high Mach number.  The shock is located near
the southern boundary of the radio halo.  The complex temperature
structure is confirmed by the more sensitive data published by Govoni
et al. (2004).  The low brightness cluster X-ray emission is elongated
toward the NW, which is the apparent direction of the merger with a
smaller subcluster.  This is the same direction as the radio halo
emission.

{\bf A~2163}.  This cluster is one of the hottest and most X-ray
luminous among known clusters.  Herbig \& Birkinshaw (1994) first
reported the presence of a powerful radio halo, which has been studied
in better detail by Feretti et al. (2001). The radio halo, which is one
of the most powerful and extended halos known so far, displays a
regular shape, slightly elongated in the E-W direction.  In the N-E
peripheral cluster region, a diffuse elongated emission, classified as
a cluster relic, has been detected.  The comparison between the radio
emission of the halo and the cluster X-ray emission shows a close
structural similarity on the large scale (Feretti et al. 2001).

According to the X-ray morphological study of Elbaz \etal (1995),
based on ROSAT data, and the spectroscopic analysis of Markevitch
\etal (1996), based on ASCA data, the cluster is likely to be a recent
merger.  The $\beta$ model gives a good fit to the average X-ray
surface brightness profile ($\chi^2$ = 49 for 36 dof), although the
cluster is not nicely spherically symmetric and the results obtained
depend on the exact region chosen to derive the profile.

Recent Chandra data (Markevitch \& Vikhlinin 2001 and Govoni et
al. 2004) reveal a complex morphology indicating that the cluster
central region is in a state of violent motion.  The temperature map
is also complex, with variations by at least a factor of 2, suggesting
streams of hot and cold gas flowing in different directions, as well
as a possible remnant of a cool gas core, surrounded by shock-heated
gas.

The structure in the temperature map is too complicated to easily
infer the geometry of the merger. The X-ray brightness distribution
indicates elongation in the E-W direction, suggesting that this could
be the merger direction. However, the absence of any sharp features in
the X-ray image, and the optical spectroscopic data (Soucail et al.,
in preparation) point to the possibility that the merger is occurring
at a large angle to the sky plane.

A strong Sunyaev-Zel'dovich (SZ) effect has been reported in this
cluster (Holzapfel \etal 1997, D\'esert \etal 1998).

\section{Radio Observations}

Radio observations were obtained with the VLA at 321.5 and 327.5 MHz
in several different configurations (see Table 2).

Observations were performed in spectral line mode, with 32 channels,
to reduce the effects of bandwidth smearing and allow for more
accurate excision of narrow-band external RFI.  For both targets, the
bandpass calibration was performed from observations of Cygnus A,
whereas 3C286 was used for initial amplitude and phase calibration and
to set the VLA flux density scale.  The sources 0834+555 and 1605-175
were observed as secondary phase calibrators for A665 and A2163,
respectively.  The entire field was imaged using the AIPS program
IMAGR in 3D mode and selfcalibrated using CALIB.  This procedure was
performed using the task VLALB, kindly provided by W. Cotton.  Further
data editing, loops of selfcalibration and wide-field image
deconvolution were used to mitigate confusion and achieve the maximum
sensitivity.

For A2163, the data from the B and C configurations have been first
reduced and self-calibrated separately, then combined to produce the
final image.  For A665, the radio halo is not detected in the B array
observation, whereas the data from the D configuration are too noisy
to be useful.  Therefore the final image for this cluster is
essentially obtained with the C configuration data.

The achieved sensitivities are 0.9 mJy/beam for A665 and 0.4 mJy/beam
for A2163.  These are somewhat higher than the theoretically expected
thermal noise levels ($\sim$ 0.55 mJy/beam in 1 hour integration),
probably because of confusion and of broad-band radio frequency
interference, mainly VLA-generated.

\section{Results}

The final images of the radio halos in A665 and A2163 at 90 cm are
presented in Figs. 1 and 2, respectively.

In A665, the radio emission is rather centrally condensed, with low
brightness features extending to the south and to the north.  The
extreme narrow region of the southern extension is due to the blend of
discrete sources, as deduced by the comparison with the 20 cm image.
Therefore the diffuse halo emission is asymmetric toward the N-NW
direction. The X-ray gas distribution is asymmetric in the same
direction (see e.g. the Chandra image by Markevitch \& Vikhlinin
2001).  The total flux of the halo, excluding the unrelated sources,
is 197 $\pm$ 6 mJy. This value leads to a total spectral index
$\alpha_{0.3}^{1.4}$ = 1.04 $\pm$ 0.02.  The central strongest halo
region has a quite regular shape, with a FWHM of about 180\arcsec,
corresponding to $\sim$ 700 kpc.  We note that this is the cluster region
approximately within 1 core radius (see Tab. 1).

In A2163, the structure of the diffuse emission is remarkably similar
to that detected at 1.4 GHz, with a similar extent.  The total flux,
after subtraction of discrete sources is 861 $\pm$ 10 mJy. This leads
to a total spectral index $\alpha_{0.3}^{1.4}$ = 1.18 $\pm$ 0.04.
Feretti et al. (2001) estimated a spectral index of $\sim$ 1.6 $\pm$
0.3 by comparing the total flux densities at the two nearby
frequencies of 1.365 MHz and 1.465 MHz within the 20 cm radio band.
Even allowing for the large errors, this could be indication of a
spectral steepening above $\sim$ 1.4 GHz.

The extended feature that was classified as a relic by Feretti et
al. (2001) is indicated in Fig. 2 by the letter ``R''. It has a total
flux of 82 $\pm$ 2 mJy, and therefore a total spectral index
$\alpha_{0.3}^{1.4}$ = 1.02 $\pm$ 0.04.

Feretti et al. (2001) also discussed the possibility that this feature
may be part of a wide-angle-tailed (WAT) radio source with the nucleus
coincident with the strong radio source J1616-061, at position RA =
16$^{\rm h}$ 16$^{\rm m}$ 22.2$^{\rm s}$, DEC = --06\degrees~
06\arcmin~34\arcsec (i.e. to the S-E of the relic).  Arguments against
this interpretation were that the WAT radio source would have the
exceptionally large size of about 1.3 Mpc (assuming it is a cluster
radio galaxy) and that WAT sources are usually located at the cluster
centers.  The present observations lend support to the interpretation
that this feature is a relic.  In fact, the strong radio source
J1616-061 seems disconnected from the source ``R''.  More important,
we derive that the spectrum of source ``R'' is flatter at its peak
location, and shows slight steepening toward the edges (see Fig. 4),
whereas a connection to the source J1616-061 would imply a progressive
spectral steepening from its nucleus to the end of the tail.  From the
morphological and spectral trend, the source J1616-061 is likely to be
a tailed radio galaxy, extended toward the south.

\section{Spectral index maps}

The spectral index maps of A665 and A2163 were obtained by comparing
the 90 cm and 20 cm images, produced with the same beam and cellsize.
The data at 90 cm provide a better coverage of short spacings than
those at 20 cm.  Indeed the minimum spacing at 90 cm is $\sim$ 35
$\lambda$, whereas at 20 cm it is $\sim$ 150 $\lambda$.  However we
are confident that the spectral comparison is not affected for the
following reasons: i) the total structures of the halos in A665 and
A2163 are extended about 10\arcmin~ and about 11.5\arcmin,
respectively.  Therefore, the structure of both halos is expected to
be fully imaged from the 20 cm observations, which are blind to
structures larger than $\sim$ 15\arcmin; ii) the shortest baselines at
90 cm are provided by a few interferometers only (C array), so the
coverage of short spacings is poor at 90 cm; iii) the total spectral
index between 90 and 20 cm is not very steep ($\alpha \sim $ 1). If
some flux was missed in the 20 cm data the real spectrum should be
flatter, in contrast to values of spectra of radio halos; iv) the 20
cm map is much more sensitive than that at 90 cm.

{\bf A~665}. The spectral index map is clumpy (top panel of
Fig. 3). The spectrum in the central halo region is rather constant,
with spectral index values between 0.8 and 1.2 within one core radius
from the cluster center (i.e. within $\sim$ 95\arcsec).  In the
northern region of lower radio brightness, the spectrum is flatter
than in the southern halo region.

Starting from the approximate radio peak position, we obtained
profiles of the spectral index trend by averaging the values of the
spectral index within small sectors around the two directions marked
by dashed lines.  The spectrum in the N-W direction flattens up to a
distance of about 200\arcsec~ from the center (Fig. 3, bottom left
panel, green triangles).  This is the region where asymmetric extended
X-ray emission is present, indicating the existence of an ongoing
merger with another cluster.  The gas temperature in this region
(Markevitch \& Vikhlinin 2001, Govoni et al. 2004) shows strong
variations, from about 12 keV in the N-E to about 8 keV in the
S-W. The spectral index flattening follows the X-ray morphology, but
there is no one-to-one correspondence with the value of the gas
temperature.  Therefore, it seems that this region, which is presently
strongly influenced by the merger, is a shocked region, where the gas
at different temperatures is still in the process of mixing.

In defining a profile in the southern cluster region, we tried to
avoid the region where there is a possible contamination of discrete
sources.  The spectrum in this direction (Fig. 3, bottom left panel,
red dots) steepens significantly from the center to the periphery.
The spectral index  increase from $\alpha$ $\sim$ 1 to $\alpha$ \gtsim~
2 is gradual and rather regular, and occurs on a scale of less than 3
cluster core radii.  The region of constant spectral index at distance
between 120\arcsec~ and 220\arcsec~ from the center is located NE of
the discrete source and could be affected by the presence of this
source.  We note that this path crosses the region of the hot shock
detected by Chandra at the southernmost edge of the radio halo
(Markevitch \& Vikhlinin 2001). The shock is located at about
100\arcsec~ from the center along the profile. No significant spectral
flattening is detected at this position.

{\bf A~2163}. The spectral index map, shown in the top panel of
Fig. 4, is clumpy in the central cluster region, but it is rather
constant within one core radius (corresponding to $\sim$ 70\arcsec),
showing spectral index values between 1 and 1.1.  On a larger scale,
there is evidence that the western halo region is flatter than the
eastern one. Particularly, there is a vertical region crossing
the cluster center and showing flatter spectrum, with a clear evidence
of spectral flattening both at the northern and at the southern halo
boundaries.

Radial profiles of the spectral index along two interesting directions
(see dashed lines) have been obtained, as in A665, by averaging the
values of the spectral index within small sectors around the two
directions.  The spectral index profile along the N-S direction
(Fig. 4, bottom left panel, green triangles) is globally rather flat
and shows a significant flattening at about 300\arcsec~ from the
center (note that the strong flattening of the two last points is due
to the presence of a discrete radio source).  In the eastern cluster
region, in the area free of discrete sources, the spectrum becomes
progressively steeper from the center to the periphery. This is well
seen in the profile along the S-E direction, given by red dots in the
left bottom panel of Fig. 4.  The spectral steepening is weaker than
in A665, and occurs over a much larger scale.  The north-south region
of flatter spectrum coincides with the region of highest temperature
detected from Chandra.  Although the geometry of the merger cannot be
established by the X-ray Chandra data (Markevitch \& Vikhlinin 2001,
Govoni et al. 2004), the region of flatter spectrum is likely related
to the strong dynamical activity at the cluster center.  The N-S
extent of the region with flat spectrum is in support of a merger
occurring in the E-W direction, as indicated by the X-ray brightness
distribution. The complexity of the merger is, however, reflected in
the complexity of the spectral index map.

\section{Discussion}

\subsection{Spectral index behaviour}

The maps reported in Figs. 3 and 4 indicate the existence of patches
of different spectral index values, with significant variations on
scales of the order of the observing beam ($\sim$ 200 kpc).  This
suggests a complex shape of the electron spectrum, as generally
expected in the case of particle reacceleration.

The regions currently influenced by an ongoing merger show a different
behaviour with respect to the relatively undisturbed regions.  Regions
of flatter spectra are indication of the presence of more energetic
radiating particles, and/or of a larger value of the local magnetic
field strength.  As qualitatively expected by electron reacceleration
models, relatively flatter regions of synchrotron spectral index are
found in regions currently influenced by merger processes, while a
general radial spectral steepening is found in presently undisturbed
cluster regions.

Numerical simulations show that the intensity of the cluster magnetic
field can be significantly enhanced during ongoing mergers (Roettiger
et al. 1999).  In addition mergers are believed to be responsible for
the injection of cluster turbulence (Norman \& Bryan 1999) which is
most likely connected to the acceleration of relativistic electrons
(Brunetti et al. 2001; Fujita et al. 2003).  Thus, in general, in the
regions directly influenced by ongoing mergers the synchrotron
spectrum may become flatter due to an increase of both the magnetic
field strength and the acceleration efficiency.

Indeed the regions with flatter spectra are observed along profiles
which appear to trace the geometry of recent merger activity as
suggested by X-ray maps.  Our results prove that the radio spectral
index can be a powerful tracer of the current physical properties of
clusters, and confirms the importance of cluster merger in the
energetics of relativistic particles responsible for the halo radio
emission.  On the other hand, the spectral index steepens
progressively with the distance from the cluster center in the region
not presently affected by the merging processes.  This is another
indication that the energy of relativistic particles is sensitive to
the effect of mergers.

It is worth noticing that there is no evidence of spectral flattening
at the location of the hot shock detected in A665 (Markevitch \&
Vikhlinin 2001).  This is consistent with the fact that shocks in
major mergers are too weak for particle acceleration (Gabici \& Blasi
2003) and indeed the Mach number of the shock in A665 is $\sim$ 2
(Markevitch \& Vikhlinin 2001). Our result supports the scenario that
cluster turbulence might be the major responsible for the supply of
energy to the radiating electrons.

We have computed the physical conditions within the radio halos in
A665 and A2163, using standard formulae (Pacholczyk 1970) with the
standard assumptions of equal energy density in protons and electrons,
a magnetic field filling factor of 1, a spectrum extending from 10 MHz
to 100 GHz with the integrated spectral index obtained in Sect. 4.
The minimum energy density in A665 is u$_{min}$ = 2.8 10$^{-14}$ erg
cm $^{-3}$, and the corresponding equipartition magnetic field is
B$_{eq}$ = 5.5 10$^{-7}$ G.  In A2163, we obtain u$_{min}$ = 3.8
10$^{-14}$ erg cm $^{-3}$, and an equipartition magnetic field
B$_{eq}$ = 6.4 10$^{-7}$ G.

We have attempted the evaluation of the energy supplied to the halos
in the regions of flatter spectral index, by matching the
reacceleration gains and the radiative losses of the radio emitting
electrons.  In regions of identical volume and same brightness at 0.3
GHz, a flattening of the spectral index from 1.3 to 0.8 implies that
the energy injected into the electron population is larger by a factor of
$\sim$2.5.

Alternatively, if electrons have been reaccelerated in the past, and
they are simply ageing, we can consider that the flatter spectrum
reflects a spectral cutoff at lower energies.  Particles with a
frequency cutoff at $\nu_b$ = 1.4 GHz have a lifetime of about 4.5
10$^7$ yr, whereas for $\nu_b$ = 0.3 GHz the particle lifetime is 9.4
10$^7$ yr, assuming a magnetic field of about 0.5 $\mu$G. Therefore it
follows that the electrons in the flat spectrum regions have been
reaccelerated more recently.

\subsection{Implications on particle acceleration and
cluster magnetic field}

In the primary reacceleration scenario the electrons are accelerated
up to a maximum energy which is given by the balance between
acceleration efficiency and energy losses. As a consequence a break
(or cut-off) is expected in the synchrotron spectrum emitted by these
electrons.  The presence of spectral steepenings and flattenings
implies that the frequency of such a synchrotron break is relatively
close to the range 0.3--1.4 GHz for a relevant fraction of the cluster
volume.

If Fermi-like acceleration processes are efficient in the cluster
volume, the synchrotron break frequency $\nu_{\rm b}$ is related to the
magnetic field B as:

\begin{equation}
\nu_{\rm b}(r) \propto \chi^2(r) 
\cases{
B(r), & if $B(r) << B_{IC}$; \cr
                             \cr
B^{-3}(r), & if $B(r) >> B_{IC}$. \cr}
\end{equation}

\noindent
where $\chi(r)= \tau_{acc}^{-1}$ is the acceleration 
rate ($dp/dt = \chi p$), and $B_{IC}$ is the IC equivalent 
magnetic field $\sim 3 \mu$G.

The equipartition magnetic field evaluated in the previous section is
$<< B_{IC}$ in the present clusters, thus we expect $\nu_{\rm b}(r)
\propto \chi^2(r)B(r)$.  We remark that this value is averaged on the
radio halo volume, therefore it represents the large scale field.
Allowing for a decrease of $B$ with $r$, it follows that a roughly
constant acceleration efficiency results in a systematic steepening of
the synchrotron spectrum with $r$, simply because at a given frequency
higher energy electrons emit in the lower field intensity.  This
steepening effect would be further enhanced if the reacceleration
efficiency increases toward the central regions.

In the bottom right panels of Figs. 3 and 4, we report the trends of
$B \chi^2$ for A665 and A2163, respectively, as a function of $r$, for
the undisturbed regions.  Each profile, normalized to its value at the
cluster center, reflects the trend of the spectral index (shown by red
dots in the bottom left panels of the same figures).  To obtain the
radial dependence of the break frequency (and thus $B \chi^2$) we have
fitted the radial behaviour of the synchrotron spectral index with
synchrotron spectra emitted by a population of accelerated electrons.
In the calculations of the electron spectra we have assumed that the
electrons have been accelerated at the same time. To constrain the
magnetic field intensity, we have used a range of electron spectra
obtained by assuming different strengths of the acceleration
coefficient and different duration times of the acceleration phase.
We have also included the uncertainties due to the deprojection of the
synchrotron spectra with distance.

In each cluster, the trend of $B \chi^2$ represents the profile of the
magnetic field strength, in the case that the reacceleration is
constant throughout the cluster.  Under this hypothesis, we can
compare the profile of the magnetic field with that obtained in the
case of a frozen-in magnetic field, $B \propto n_{th}^{2/3}$, where
$n_{th}$ is the thermal plasma density.  This profile is expected if
the intensity of the magnetic field results from the compression of
the thermal plasma during the cluster gravitational collapse.  We
derive that this trend is not a good representation of the
observations, expecially in the case of A2163. Indeed, the intensity
of the frozen-in magnetic field decreases by 50\% at 130\arcsec~ for
A665, and at 110\arcsec~ for A2163, much steeper than estimated by the
spectral behaviour.  We note, however, that in A665 and A2163 the
$\beta$ model approximation for the distribution of $n_{th}$ may be a
too rough representation of the gas distribution, owing to the presence of
gas perturbations as a result of strong dynamical evolution.  On
the other hand, detailed MHD numerical simulations show that the
radial behaviour of the magnetic field with radius may diverge from
the prediction of a frozen-in B model, resulting flatter in the
central regions, and steeper in the external regions (Dolag et
al. 1999, 2002).

Radial profiles of the cluster magnetic fields have been derived in
the literature for Coma by modeling the spectral steepening (Brunetti
et al. 2001) and for A119, using the Rotation Measure of cluster radio
galaxies (Dolag et al. 2001).  In both clusters, the magnetic field
profiles decline with the distance and are not too different from
those expected by the frozen-in B model.

We finally remark that the hypothesis of constant reacceleration
efficiency in the cluster volume may not be valid, according to the
following arguments; i) first results from numerical simulations
(Norman \& Bryan 1999) indicate that the injection of turbulence in
clusters is not homogeneous and occurs on very different scales in
different regions; ii) detailed calculations of particle acceleration
due to Alfv\`en waves show that, under reasonable assumptions, the
acceleration efficiency slightly increases with distance from the
cluster center (Brunetti et al. 2003); iii) the radiative losses of
electrons in the innermost cluster regions may be strongly increased
if the magnetic field is larger than the equipartition value, in
particular if it is of the order of B$_{IC}$ (e.g Kuo et al. 2003).

In general, the ongoing violent mergers in A665 and A2163 are likely
to play a crucial role in determining the conditions of the radiating
particles and the magnetic field in these clusters.

\section{Conclusions}

In this paper we have presented new images at 0.3 GHz of the radio
halos in the clusters A665 and A2163 and we have derived the spectral
index maps between 0.3 and 1.4 GHz, using maps already published at
1.4 GHz.

Our results show that the integrated radio spectra of these halos are
moderately steep in this frequency range: $\alpha^{1.4}_{0.3}$ = 1.04
and $\alpha^{1.4}_{0.3}$ = 1.18, for A665 and A2163, respectively.

The spectral index maps, produced with an angular resolution of the
order of $\sim$ 1\arcmin, show a clumpy distribution with significant
variations, which are indication of a complex shape of the radiating
electron spectrum, and are therefore in support of halo models invoking the 
reacceleration of relativistic particles.

We find that the regions of flatter spectrum appear to trace the
geometry of recent merger activity as suggested by X-ray maps.  These
results prove that the radio spectral index can be a powerful tracer
of both the current physical properties and past history of clusters.

We find no evidence of spectral flattening at the location of the hot
shock detected in A665 (Markevitch \& Vikhlinin 2001).  This favours
the scenario that cluster turbulence might be the major responsible
for the electron reacceleration.

In the undisturbed cluster regions, the spectrum steepens with the
distance from the cluster center.  Brunetti et al. (2001) explain the
spectral steepening in the framework of electron reacceleration
models, as due to the combined effect of a radial decrease of the
cluster magnetic field strength and of the presence of a high energy
break in the energy distribution of the reaccelerated electron
population. This is more difficult to reproduce by other models, such
as those requiring secondary electron populations.

The spectral steepening detected in A665 and A2163 in the direction not 
presently affected by the merging processes allowed us to constrain
the profile of the product between the cluster magnetic field
and the reacceleration efficiency, $B \chi^2$, under simple assumptions.

The magnetic field profile in both clusters is flatter than predicted
in the case of a constant reacceleration and magnetic field frozen to
the cluster thermal gas.

The ongoing violent mergers may play a crucial role in determining the
conditions of the radiating particles and of the magnetic fields in
clusters.

\begin{acknowledgements}

The National Radio Astronomy Observatory is operated by Associated
Universities, Inc., under contract with the National Science
Foundation.  Basic research in radio astronomy at the Naval Research
Laboratory is supported by the Office of Naval Research.  This work
has been partially funded by the Italian Space Agency (ASI).

\end{acknowledgements}

\begin{figure*}
\vspace{3 cm}
\caption{{\bf Top panel}: color-scale image of the spectral index
between 0.3 GHz and 1.4 GHz of A665, obtained with a resolution of
68\arcsec $\times$ 59\arcsec (PA= 25\degrees) FWHM.  The contours
indicate the radio emission at 20 cm (from Giovannini \& Feretti 2000)
at levels of -0.2, 0.2, 0.4, 0.8, 1.5, 3.0, 6.0, 12.0, 25.0 mJy/beam
(beam = 52\arcsec $\times$ 42\arcsec).  {\bf Bottom left panel}:
radial profiles of the spectral index along the two directions
indicated by the dashed lines in the spectral index map: the green
triangles report the profile toward the N-W (green dashed line), the
red dots give the profile toward the S-E (red dashed line).  In both
profiles, the origin of the distance scale is the approximate radio
peak position at the cluster center (RA$_{2000}$ = 08$^h$ 30$^m$
56.0$^s$, DEC$_{2000}$ = 65\degrees~ 50\arcmin~ 09.7\arcsec).  {\bf
Bottom right panel}: radial profile of the product between the
magnetic field and the electron reacceleration coefficient, normalized
to its value at the custer center}
\end{figure*}

\begin{figure*}
\vspace{3 cm}
\caption{ {\bf Top panel}: color-scale image of the spectral index
between 0.3 GHz and 1.4 GHz of A2163, obtained with a resolution of
60\arcsec $\times$ 51\arcsec (PA=0\degrees) FWHM.  The contours
indicate the radio emission at 20 cm (from Feretti et al. 2001) at
levels of 0.2, 0.5, 0.8, 1.5, 3.0, 5.0, 7.0, 9.0, 15.0, 25.0 mJy/beam
(beam = 60\arcsec $\times$ 45\arcsec).  {\bf Bottom left panel}:
radial profiles of the spectral index along the two directions
indicated by the dashed lines in the spectral index map: the green
triangles report the profile toward the N (green dashed line), the red
dots give the profile toward the S-E (red dashed line).  In both
profiles, the origin of the distance scale is the approximate center
of the radio emission (RA$_{2000}$ = 16$^h$ 15$^m$ 51.2$^s$,
--06\degrees~ 08\arcmin~ 48.1\arcsec, note that the radio peak to the
east is contaminated by discrete sources). {\bf Bottom right panel}:
radial profile of the product between the magnetic field and the
electron reacceleration coefficient, normalized to its value at the
cluster center.}
\end{figure*}

\end{document}